\documentclass[11pt,fleqn]{article}

\usepackage{amsmath,amssymb,amsthm} 

\newcommand{\p}{{\partial}}
\newcommand{\rank}{\mathop{\rm rank}\nolimits}

\newcommand{\Int}{\mathop{\rm Int}\nolimits}

\newcommand{\diag}{\mathop{\rm diag}\nolimits}

\newtheorem{theorem}{Theorem}

\newtheorem{lemma}{Lemma}
{\theoremstyle{definition}

\newtheorem{note}{Note}
\newtheorem*{note*}{Note}

}

\begin{document}

\allowdisplaybreaks

\begin{flushleft}
\LARGE \bf Invariants of Lie Algebras
\\ with Fixed Structure of Nilradicals
\end{flushleft}

\begin{flushleft} \bf
Vyacheslav Boyko~$^\dag$, Jiri Patera~$^\ddag$ and Roman Popovych~$^{\dag\S}$
\end{flushleft}

\noindent $^\dag$~Institute of Mathematics of NAS of Ukraine, 3
Tereshchenkivs'ka Str., Kyiv-4, 01601 Ukraine\\
$\phantom{^\dag}$~E-mail: boyko@imath.kiev.ua, rop@imath.kiev.ua

\noindent
$^\ddag$~Centre de Recherches Math\'ematiques,
Universit\'e de Montr\'eal,\\
$\phantom{^\ddag}$~C.P. 6128 succursale Centre-ville, Montr\'eal (Qu\'ebec), H3C 3J7 Canada\\
$\phantom{^\ddag}$~E-mail: patera@CRM.UMontreal.CA

\noindent $^\S$~Fakult\"at f\"ur Mathematik, Universit\"at Wien, Nordbergstra{\ss}e 15, A-1090 Wien, Austria

\begin{abstract}
\noindent
An algebraic algorithm is developed for computation of invariants
(`generalized Casimir operators') of general Lie algebras over the
real or complex number field. Its main tools are the Cartan's
method of moving frames and the knowledge of the group of inner
automorphisms of each Lie algebra. Unlike the first application of
the algorithm in [{\it J.\,Phys.\,A: Math.\,Gen.}, 2006, V.39, 5749; math-ph/0602046], 
which deals with low-dimensional Lie algebras, 
here the effectiveness of the algorithm is demonstrated by
its application to computation of invariants of  solvable Lie
algebras of general dimension $n<\infty$ restricted only by a
required structure of the nilradical.

Specifically, invariants are calculated here for 
families of real/complex solvable Lie algebras. These families contain,
with only a few exceptions, all the solvable Lie algebras of
specific dimensions, for whom the invariants are found in the
literature.
\end{abstract}

\section{Introduction}

The term Casimir operator was born in the physics literature about
half a century ago as a reference to  \cite{Casimir}.  At that time
only the lowest rank Lie algebras appeared of interest. In
subsequent years the need to know the invariant operators of much
larger Lie algebras grew more rapidly in physics than in
mathematics.

In the mathematics literature it was soon recognized that the
universal enveloping algebra $U(\mathfrak{g})$ of a semisimple Lie algebra $\mathfrak{g}$
contains elements that commute with $\mathfrak{g}$, that there is a basis for
all such invariants and that the number of basis elements
coincides with the rank of $\mathfrak{g}$. The degrees of the basis
elements are given by the values of the exponents of the
corresponding Weyl group (augmented by~$1$). The exponents are
listed in many reference texts, see for example \cite{BMP}. Best known
are the Casimir operators of degree $2$ for every semisimple Lie
algebra. The actual form of a Casimir operators dependents on the
choice of basis of $\mathfrak{g}$.

Soon after the analogous question about the invariant operators 
was asked also for non-semi\-simple Lie algebras. An answer
exhausting all cases appears out of reach at present. However
methods applicable to specific Lie algebras were invented and used~\cite{Patera&Sharp&Winternitz1976}.

There are numerous papers on properties and specific computation 
of invariant operators of Lie algebras, on estimation of their
number and on application of invariants of various classes of
Lie algebras,  or even a particular Lie algebra which appears in
physical problems (see~\cite{Ancochea&Campoamor-Stursberg&GarciaVergnolle2006,Boyko&Patera&Popovych2006,%
Campoamor-Stursberg2005b,Campoamor-Stursberg2006a,%
Ndogmo2000a,Ndogmo&Wintenitz1994a,Ndogmo&Wintenitz1994b,Patera&Sharp&Winternitz1976,
Rubin&Winternitz1993,Snobl&Winternitz2005,Tremblay&Winternitz1998,Tremblay&Winternitz2001} 
and references therein).

  The purpose of the paper is to present the latest version of the
method first used for low-dimensional Lie algebras in \cite{Boyko&Patera&Popovych2006},
and to demonstrate its effectiveness by computing the invariants for
families of solvable Lie algebras of general
dimension. The families are distinguished by the structure of the
nilradicals of their Lie algebras.

The main advantage of the method is in that it is purely algebraic. It eliminates
the need to solve systems of differential equations of the
conventional method, replacing them by algebraic equations.
Moreover, efficient exploitation of the new method imposes certain
constrains on the choice of bases of the Lie algebras. This then
leads to simpler expressions for the invariants. In some cases the
simplification is considerable.

Our paper is organized as follows. 

After short review of necessary notions and results in Section~\ref{SectionPreliminaries}, we formulate
the algebraic algorithm of construction  of the generalized Casimir
operators of Lie algebras (Section~\ref{SectionAlgorithm}). It is based on approach introduced in
\cite{Boyko&Patera&Popovych2006} for the case of algebras of arbitrary (fixed)
dimension.  The algorithm makes use of the Cartan's method of
moving frames in the Fels--Olver version~\cite{Fels&Olver1998,Fels&Olver1999}. 
More exactly, the notion of lifted invariants and different techniques of 
excluding parameters are applied.

In Section~\ref{SectionIllustrativeExample} an illustrative example on invariants of 
a six-dimensional algebras is given for clear demonstration of features of the developed method.
The main subject of our interest in the present paper are invariants, 
generalized Casimir operators, of solvable Lie algebras of arbitrary finite dimension $n<\infty$. 
For convenience all necessary notations are collected in separate Section~\ref{SectionNotations}.
A number of families of Lie algebras are considered further. 
The families are distinguished by the structure of their nilradicals. 
The invariant operators are found at once for all members of the family. 

The Lie algebras with Abelian ideals of codimension 1 are completely investigated 
in the case of the both complex and real fields in Section~\ref{SectionOnAlgsWithAbIdialsOfCodim1}. 
The nilradicals of the algebras studied in Section~\ref{SectionOnAlgsWithSimplestFiliformNilradicals} are 
isomorphic to the simplest filiform algebras. 
Consideration of nilpotent algebras of strictly upper triangle matrices 
in Section~\ref{SectionOnAlgebraOfStrictlyUpperTriangleMatrices} is most sophisticated. 
At the same time, the developed method allows us to clarify an origin of the Casimir operators for 
these algebras, which was first found in~\cite{Tremblay&Winternitz2001}.

All these examples illustrate various aspects and advantages of the proposed method. 

\section{Preliminaries}\label{SectionPreliminaries}

Consider a Lie algebra~$\mathfrak{g}$ of dimension $\dim \mathfrak{g}=n<\infty$ over the complex or real field 
and the corresponding connected Lie group~$G$.
Let~$\mathfrak{g}^*$ be  the dual space of the vector space~$\mathfrak{g}$.
The map ${\rm Ad}^*\colon G\to GL(\mathfrak{g}^*)$ defined for any $g\in G$ by the relation
\[
\langle{\rm Ad}^*_g x,u\rangle=\langle x,{\rm Ad}_{g^{-1}}u\rangle
\quad \mbox{for all $x\in \mathfrak{g}^*$ and $u\in \mathfrak{g}$}
\]
is called the {\it coadjoint representation} of the Lie group~$G$. 
Here ${\rm Ad}\colon G\to GL(\mathfrak{g})$ is the usual adjoint representation of~$G$ in~$\mathfrak{g}$,
and the image~${\rm Ad}_G$ of~$G$ under~${\rm Ad}$ is the inner automorphism group ${\rm Int}(\mathfrak{g})$ 
of the Lie algebra~$\mathfrak{g}$.
The image of~$G$ under~${\rm Ad}^*$ is a subgroup of~$GL(\mathfrak{g}^*)$ and is denoted by~${\rm Ad}^*_G$. 

A function $F\in C^\infty(\mathfrak{g}^*)$ is called an {\it invariant} of~${\rm Ad}^*_G$
if $F({\rm Ad}_g^* x)=F(x)$ for all $g\in G$ and $x\in \mathfrak{g}^*$.

The set of invariants of ${\rm Ad}^*_G$ is denoted by $\mathop{\rm Inv}\nolimits({\rm Ad}^*_G)$.
The maximal number $N_\mathfrak{g}$ of functionally independent invariants in $\mathop{\rm Inv}\nolimits({\rm Ad}^*_G)$
coincides with the codimension of the regular orbits of~${\rm Ad}^*_G$, i.e. 
it is given by the difference
\[
N_\mathfrak{g}=\dim \mathfrak{g}-\rank {\rm Ad}^*_G.
\]
Here $\rank {\rm Ad}^*_G$ denotes the dimension of the regular orbits of~${\rm Ad}^*_G$. 
It is a basis independent characteristic of the algebra~$\mathfrak{g}$, 
the same as $\dim \mathfrak{g}$ and $N_\mathfrak{g}$.
Sometimes $\rank {\rm Ad}^*_G$ is called as the rank of the Lie algebra~$\mathfrak{g}$ or the Dixmier's invariant.
(Let us note that the first name is more often used for other numerical characteristics of Lie algebras, 
which can differ from the above one~\cite{Jacobson}.)

To calculate invariants explicitly, one should fix a basis of the algebra. 
Any (fixed) set of basis elements $e_1,\ldots,e_n$ of~$\mathfrak{g}$ satisfies the commutation relations
\begin{gather*}
[e_i,e_j]=c_{ij}^k e_k,\quad i,j,k=1,\ldots,n,
\end{gather*}
where $c_{ij}^k$ are components of the tensor of structure constants of~$\mathfrak{g}$ in the chosen basis.

Let $x\to\check x=(x_1,\ldots,x_n)$ be the coordinates in~$\mathfrak{g}^*$ associated with 
the dual basis to the basis $e_1,\ldots, e_n$.
Given any invariant $F(x_1,\ldots,x_n)$ of~${\rm Ad}^*_G$, one finds
the corresponding invariant of the Lie algebra~$\mathfrak{g}$ as symmetrization, 
$\mathop{\rm Sym}\nolimits F(e_1,\ldots,e_n)$, of $F$. 
It is often called a \emph{generalized Casimir operator} of~$\mathfrak{g}$.  
If $F$ is a~polynomial, $\mathop{\rm Sym}\nolimits F(e_1,\ldots,e_n)$ is a usual \emph{Casimir operator}, 
i.e. an element of the center of the universal enveloping algebra of~$\mathfrak{g}$.
More precisely, the symmetrization operator~$\mathop{\rm Sym}\nolimits$ acts only on the monomials
of the forms~$e_{i_1}\cdots e_{i_r}$,
where there are non-commuting elements among~$e_{i_1}, \ldots, e_{i_r}$, and is defined by the formula
\[
\mathop{\rm Sym}\nolimits (e_{i_1}\cdots e_{i_r})=\dfrac1{r!}\sum_{\sigma\in S_r}
e_{i_{\sigma_1}}\cdots e_{i_{\sigma_r}},
\]
where $i_1, \ldots, i_r$ take values from 1 to $n$, $r\in\mathbb{N}$,
the symbol $S_r$ denotes the permutation group of $r$ elements.
The set of invariants of $\mathfrak{g}$ is denoted by $\mathop{\rm Inv}\nolimits(\mathfrak{g})$.

A set of functionally independent invariants
$F^l(x_1,\ldots,x_n)$, \mbox{$l=1,\ldots,N_\mathfrak{g}$},
forms {\it a~functional basis} ({\it fundamental invariant})
 of $\mathop{\rm Inv}\nolimits({\rm Ad}^*_G)$, i.e.\
any invariant $F(x_1,\ldots,x_n)$ can be uniquely presented as a~function 
of~$F^l(x_1,\ldots,x_n)$, \mbox{$l=1,\ldots,N_\mathfrak{g}$}.
Accordingly the set of $\mathop{\rm Sym}\nolimits F^l(e_1,\ldots,e_n)$, \mbox{$l=1,\ldots,N_\mathfrak{g}$}, 
is called a basis of~$\mathop{\rm Inv}\nolimits(\mathfrak{g})$.

If the Lie algebra $\mathfrak{g}$ is decomposable into 
the direct sum of Lie algebras~$\mathfrak{g}_1$ and~$\mathfrak{g}_2$ 
then the union of bases of~$\mathop{\rm Inv}\nolimits(\mathfrak{g}_1)$ 
and~$\mathop{\rm Inv}\nolimits(\mathfrak{g}_2)$ is a basis of~$\mathop{\rm Inv}\nolimits(\mathfrak{g})$. 
Therefore, for classification of invariants of Lie algebras from a given class 
it is really enough for ones to describe only invariants of the indecomposable algebras from this class.

Our task here is to determine the basis of the functionally independent invariants for
${\rm Ad}^*_G$ and then to transform these invariants to the invariants of the algebra~$\mathfrak{g}$.
Any other invariant of $\mathfrak{g}$ is a function of the independent ones. 

Any invariant $F(x_1,\ldots,x_n)$ of~${\rm Ad}^*_G$ is a solution of
the linear system of first-order partial differential equations
\begin{gather*}
X_iF=0,\quad \mbox{i.e.}\quad c_{ij}^k x_kF_{x_j}=0,
\end{gather*}
where $X_i=c_{ij}^k x_k\partial_{x_j}$ is the infinitesimal generator
of the one-parameter group $\{{\rm Ad}^*_G(\exp\varepsilon e_i)\}$
corresponding to $e_i$. The mapping $e_i\to X_i$ gives a representation of the Lie algebra~$\mathfrak{g}$. 
It is faithful iff the center of $\mathfrak{g}$ consists of zero only. 
In the terms of structure constants for the fixed basis, 
the rank of coadjoint representation can be found by the formula
\[
\rank {\rm Ad}^*_G=\sup\limits_{\check x\in\mathbb{R}^n}\rank\, (c_{ij}^k x_k)_{i,j=1}^n.
\]

The standard method
of construction of generalized Casimir operators consists of integration 
of the above system of partial differential equations.
It turns out to be rather cumbersome calculations,
once the dimension of Lie algebra is not one of the lowest few.
Alternative methods use matrix representations of Lie algebras. 
They are not much easier and are valid for a limited class of representations.

The algebraic method of computation of invariants of Lie algebras presented 
in this paper is simpler and generally valid.
It extends to our problem the exploitation of 
the Cartan's method of moving frames~\cite{Fels&Olver1998,Fels&Olver1999}.

\section{The algorithm}\label{SectionAlgorithm}

Let us recall some facts from \cite{Fels&Olver1998,Fels&Olver1999} and adapt them
to the particular case of the coadjoint action of~$G$ on~$\mathfrak{g}^*$.
Let~$\mathcal{G}={\rm Ad}^*_G\times \mathfrak{g}^*$ denote 
the trivial left principal ${\rm Ad}^*_G$-bundle over~$\mathfrak{g}^*$.
The right regularization~$\widehat R$ of the coadjoint action of~$G$ 
on~$\mathfrak{g}^*$ is the diagonal action of~${\rm Ad}^*_G$ 
on~$\mathcal{G}={\rm Ad}^*_G\times \mathfrak{g}^*$. 
It is provided by the maps
\begin{gather*}
\widehat R_g({\rm Ad}^*_h,x)=({\rm Ad}^*_h\cdot {\rm Ad}^*_{g^{-1}},{\rm Ad}^*_g x),
\qquad g,h\in G, \quad x\in \mathfrak{g}^*,
\end{gather*}
where the action on the bundle~$\mathcal{G}={\rm Ad}^*_G\times \mathfrak{g}^*$ is regular and free. 
We call $\widehat R_g$ the \emph{lifted coadjoint action} of~$G$. 
It projects back to the coadjoint action on~$\mathfrak{g}^*$ 
via the ${\rm Ad}^*_G$-equivariant projection~$\pi_{\mathfrak{g}^*}\colon \mathcal{G}\to \mathfrak{g}^*$. 
Any \emph{lifted invariant} of~${\rm Ad}^*_G$ is a (locally defined) smooth function 
from~$\mathcal{G}$ to a~manifold, which is invariant with respect to the lifted coadjoint action of~$G$.
The function $\mathcal{I}\colon\mathcal{G}\to \mathfrak{g}^*$ 
given by $\mathcal{I}=\mathcal{I}({\rm Ad}^*_g,x)={\rm Ad}^*_g x$ 
is the \emph{fundamental lifted invariant} of ${\rm Ad}^*_G$, i.e.\ $\mathcal{I}$ is a lifted invariant and 
any lifted invariant can be locally written as a function of~$\mathcal{I}$.
Using an arbitrary function~$F(x)$ on~$\mathfrak{g}^*$, 
we can produce the lifted invariant~$F\circ\mathcal{I}$ of~${\rm Ad}^*_G$ 
by replacing $x$ with $\mathcal{I}={\rm Ad}^*_g x$ in the expression for~$F$. 
Ordinary invariants are particular cases of lifted invariants, where one identifies any invariant formed 
as its composition with the standard projection~$\pi_{\mathfrak{g}^*}$. 
Therefore, ordinary invariants are particular functional combinations
of lifted ones that happen to be independent of the group parameters of~${\rm Ad}^*_G$. 

In view of the above consideration, the proposed algorithm for the construction
of invariants of Lie algebra $\mathfrak{g}$ can be briefly formulated in the following four steps.

\medskip

1. {\it Construction of generic matrix $B(\theta)$ of~${\rm Ad}^*_G$.}
It is calculated from the structure constants of the Lie algebra by exponentiation. 
$B(\theta)$ is the matrix of an inner automorphism of the Lie algebra~$\mathfrak{g}$ in the given basis 
$e_1$, \ldots, $e_n$, $\theta=(\theta_1,\ldots,\theta_r)$ are group parameters (coordinates) 
of~$\mathop{\rm Int}(\mathfrak{g})$, and 
\begin{gather*}
r=\dim{\rm Ad}^*_G=\dim\mathop{\rm Int}(\mathfrak{g})=n-\dim{\rm Z}(\mathfrak{g}),
\end{gather*}
${\rm Z}(\mathfrak{g})$ is the center of~$\mathfrak{g}$.

\medskip

2. {\it Fundamental lifted invariant.} 
The explicit form of the fundamental lifted invariant~$\mathcal{I}=(\mathcal{I}_1,\ldots,\mathcal{I}_n)$ 
of ${\rm Ad}^*_G$ in the chosen coordinates~$(\theta,\check x)$ in ${\rm Ad}^*_G\times \mathfrak{g}^*$ is
\[
(\mathcal{I}_1,\ldots,\mathcal{I}_n)=(x_1,\ldots,x_n)\cdot B(\theta_1,\ldots,\theta_r),
\]
or briefly $\mathcal{I}=\check x\cdot B(\theta)$. 

\medskip

3. {\it Elimination of parameters by normalization}. 
We find a nonsingular submatrix 
\[
\dfrac{\p(\mathcal{I}_{j_1},\ldots,\mathcal{I}_{j_\rho})}{\p(\theta_{k_1},\ldots,\theta_{k_\rho})}
\]
of the maximal dimension~$\rho$ in the Jacobian matrix~$\p\mathcal{I}/\p\theta$ and solve the equations 
$\mathcal{I}_{j_1}=c_1$, \ldots, $\mathcal{I}_{j_\rho}=c_\rho$ with respect to 
$\theta_{k_1}$,~\ldots,~$\theta_{k_\rho}$. 
Here the constants $c_1$, \ldots, $c_\rho$ are chosen to lie in the range of values of 
$\mathcal{I}_{j_1}$, \ldots, $\mathcal{I}_{j_\rho}$. 
After substituting the found solutions into the other lifted invariants, 
we obtain $N_\mathfrak{g}=n-\rho$ usual invariants $F^l (x_1,\ldots,x_n)$.

\medskip

4. {\it Symmetrization.} The functions $F^l(x_1,\ldots,x_n)$ which form a basis 
of~$\mathop{\rm Inv}\nolimits({\rm Ad}^*_G)$ are symmetrized to
$\mathop{\rm Sym}\nolimits F^l(e_1,\ldots,e_n)$. 
It is the desired basis of~$\mathop{\rm Inv}\nolimits(\mathfrak{g})$.

\medskip

Let us give some remarks on steps of the algorithm.

In the first step we usually use second canonical coordinates on $\Int(\mathfrak{g})$ as group parameters~$\theta$
and present the matrix~$B(\theta)$ in the form  
\[
B(\theta)=\prod_{i=1}^r\exp(\theta_i\hat{\rm ad}_{e_{n-r+i}}),
\]
where $e_1$, \ldots, $e_{n-r}$ are assumed to form a basis of~$Z(\mathfrak{g})$; 
${\rm ad}_v$ denotes the adjoint representation of $v\!\in\!\mathfrak{g}$ in~$GL(\mathfrak{g})$: 
${\rm ad}_v w=[v,w]$ for all $w\!\in\!\mathfrak{g}$, 
and the matrix of ${\rm ad}_v$ in the basis $e_1$, \ldots, $e_n$ is denoted as~$\hat{{\rm ad}}_v$. 
In particular, $\hat{{\rm ad}}_{e_i}=(c_{ij}^k)_{j,k=1}^n$.
Often the parameters~$\theta$ are additionally transformed in a light manner (signs, renumbering, re-denotation etc) 
for simplification of the final presentation of~$B(\theta)$. 
It is also sometimes
 convenient for us to introduce `virtual' group parameters corresponding to center basis elements.

Since $B(\theta)$ is a general form of matrices from $\Int(\mathfrak{g})$, we should not adopt it in any way 
for the second step. 

In fact, the third step of our algorithm can involve different techniques of elimination of parameters 
which are also based on using an explicit form of lifted invariants \cite{Boyko&Patera&Popovych2006}. 
The applied normalization procedure~\cite{Fels&Olver1998,Fels&Olver1999} can also be modified and be used 
in more involved way 
(see e.g. Subsection~\ref{SubsecOnGeneralJordanForm}). 

Let us emphasize that the maximal dimension of a nonsingular submatrix in the Jacobian matrix~$\p\mathcal{I}/\p\theta$ 
coincides with the rank of coadjoint representation of~$\mathfrak{g}$, i.e.
\[
\rank {\rm Ad}^*_G=\rho=\max_{\check x\in\mathbb{R}^n}\max_{\theta\in\mathbb{R}^r}\rank \dfrac{\p\mathcal{I}}{\p\theta}.
\]
It gives one more formula for calculation of the rank of coadjoint representation.

\medskip

\looseness=-1
In conclusion let us underline that the search of invariants of Lie algebra $\mathfrak{g}$,
which has been done by solving a linear system of first-order partial differential equations,
is replaced here by the construction of the matrix~$B(\theta)$ of inner automorphisms
and by excluding the parameters~$\theta$ 
from the fundamental lifted invariant $\mathcal{I}=\check x\cdot B(\theta)$ in some way.

\section{Illustrative example}\label{SectionIllustrativeExample}
 
The six-dimensional solvable Lie algebra~$\mathfrak{g}_{6.38}^{a}$ \cite{Mubarakzyanov1963-3} 
with five-dimensional nilradical $\mathfrak{g}_{3.1}\oplus 2\mathfrak{g}_1$ 
has the following non-zero commutation relations 
\begin{gather*}
[e_4, e_5] = e_1, \quad [e_1, e_6] = 2ae_1, \quad [e_2, e_6]= a e_2- e_3,\quad [e_3, e_6] = e_2+a e_3,\\
[a_4,e_6]=e_2+ae_4-e_5, \quad [e_5,e_6]=e_3+e_4+ae_5, \quad a\in{\mathbb R}.
\end{gather*}     
Here we have modified the basis to $K$-canonical form~\cite{Mubarakzyanov1963-1}, i.e.\ now
$\langle e_1,\ldots, e_i\rangle$ is an ideal of $\langle e_1,\ldots,e_i, e_{i+1} \rangle$ for any $i=1,2,3,4,5$. 
(See also~\cite{Boyko&Patera&Popovych2006} for discussion of role of $K$-canonical bases in 
the investigation of solvable Lie algebras.) 

The matrices of the adjoint representation~$\hat{\rm ad}_{e_i}$ of the basis elements $e_1$, $e_2$, $e_3$, $e_4$, $e_5$ 
and $e_6$ correspondingly have the form
\begin{gather*}
\left(\begin{array}{cccccc} 0&0&0& 0 & 0 & 2a \\0&0&0&0&0&0\\0&0&0&0&0&0\\0&0&0&0&0&0\\0&0&0&0&0&0\\0&0&0&0&0&0\end{array}\right)\!, \quad
\left(\begin{array}{cccccc} 0&0&0&0 & 0 & 0 \\0&0&0&0 &0 & a \\0&0&0&0 &0 & -1 \\0&0&0&0 & 0&0\\0&0&0&0&0&0\\0&0&0&0&0&0\end{array}\right)\!, \quad
\left(\begin{array}{cccccc} 0&0&0&0 &0&0\\0&0&0&0 &0& 1\\0&0&0&0 & 0 & a \\0&0&0&0 &0 & 0 \\0&0&0&0&0 & 0\\0&0&0&0&0&0\end{array}\right)\!, \\
\left(\begin{array}{cccccc} 0&0&0&0 & 1 & 0 \\0&0&0&0 & 0 & 1 \\0&0&0&0 &0 & 0 \\0&0&0&0 & 0 & a \\0&0&0&0&0&-1\\0&0&0&0&0&0 \end{array}\right)\!,\quad
\left(\begin{array}{cccccc} 0&0&0&-1&0&0\\0&0&0&0&0&0 \\0&0&0&0&0&1 \\0&0&0&0&0&1 \\0&0&0&0&0&a\\0&0&0&0&0&0 \end{array}\right)\!,\quad
\left(\begin{array}{cccccc} -2a&0&0&0&0&0\\0&-a&-1&-1&0&0 \\0&1&-a&0&-1&0 \\0&0&0&-a&-1&0 \\0&0&0&1&-a&0\\0&0&0&0&0&0 \end{array}\right)\!.
\end{gather*}
The inner automorphisms of~$\mathfrak{g}_{6.38}^{a}$ are then described by the block triangular matrix 
\begin{gather*}
B(\theta)=\prod_{i=1}^5\exp(\theta_i\hat{\rm ad}_{e_i}) \cdot \exp(-\theta_6\hat{\rm ad}_{e_6})\\[1ex]
\phantom{B(\theta)}{}=\left(\begin {array}{cccccc}
\varepsilon^2 & 0 & 0 & -\theta_5\varepsilon\varkappa-\theta_4\varepsilon\sigma & -\varepsilon\theta_5\sigma+\varepsilon\theta_4\varkappa 
 & -\frac12\theta_5^2+ a\theta_4\theta_5-\frac 12\theta_4^2+2a\theta_1 \\
0&\varepsilon \varkappa&\varepsilon \sigma& \theta_6 \varepsilon \varkappa & \theta_6 \varepsilon \sigma& \theta_4+\theta_3+a\theta_2 \\
0&-\varepsilon \sigma& \varepsilon \varkappa & -\theta_6 \varepsilon \sigma &\theta_6 \varepsilon \varkappa&\theta_5+a\theta_3-\theta_2\\
0&0&0&\varepsilon \varkappa&\varepsilon \sigma&\theta_5+a\theta_4\\
0&0&0&-\varepsilon \sigma& \varepsilon \varkappa &a\theta_5-\theta_4\\
0&0&0&0&0&1
\end {array}\right),
\end{gather*}
where $\varepsilon=e^{a\theta_6}$, $\varkappa=\cos\theta_6$, $\sigma=\sin\theta_6$. 
Therefore, a functional basis of lifted invariants is formed by
\begin{gather*}
\mathcal{I}_1=\varepsilon^2 x_1,\\
\mathcal{I}_2=\varepsilon (\varkappa x_2-\sigma x_3),\\
\mathcal{I}_3=\varepsilon (\sigma x_2+\varkappa x_3),\\
\mathcal{I}_4=\varepsilon ((-\theta_5 \varkappa-\theta_4\sigma)x_1+\theta_6\varkappa x_2
-\theta_6\sigma x_3+\varkappa x_4 -\sigma x_5),\\
\mathcal{I}_5=\varepsilon ((-\theta_5 \sigma +\theta_4\varkappa)x_1+\theta_6\sigma x_2
+\theta_6\varkappa x_3+\sigma x_4 +\varkappa x_5),\\
\mathcal{I}_6=\big(-\tfrac 12\theta_5^2+ a\theta_4\theta_5-\tfrac 12\theta_4^2+2a\theta_1\big)x_1+
(\theta_4+\theta_3+a\theta_2)x_2+(\theta_5+a\theta_3-\theta_2)x_3\\
\phantom{\mathcal{I}_6=}{} +(\theta_5+a\theta_4)x_4+(a\theta_5-\theta_4)x_5+x_6.
\end{gather*}

The algebra~$\mathfrak{g}_{6.38}^{a}$ has two independent invariants. 
They can be easily found from first three lifted invariants by the normalization procedure.
Further the cases $a=0$ and $a\not=0$ should be considered separately 
since there exists difference between them in the normalization procedure. 

It is obvious in case $a=0$ that $e_1$ generating the center~$Z(\mathfrak{g}_{6.38}^{0})$ is one of the invariants.
The second invariant is found via combining the lifted invariants $\mathcal{I}_2$ and $\mathcal{I}_3$: 
$\mathcal{I}_2^2+\mathcal{I}_3^2=x_2^2+x_3^2$.
Since the symmetrization procedure is trivial for this algebra we obtain the following set of polynomial invariants
\[
e_1, \quad e_2^2+e_3^2. 
\] 

In case $a\not=0$ we solve the equation $\mathcal{I}_1=1$ with respect to $e^{2a\theta_6}$ and  
substitute the obtained expression $e^{2a\theta_6}=1/x_1$ into 
the combinations $\mathcal{I}_2^2+\mathcal{I}_3^2$ and $\exp(-2a\arctan\mathcal{I}_3/\mathcal{I}_2)$.
In view of trivial symmetrization we obtain the final basis of generalized Casimir invariants
\begin{gather*}
\frac{e_2^2+e_3^2}{e_1},\quad e_1\exp\Bigl(-2a\arctan \frac{e_3}{e_2}\Bigr). 
\end{gather*}
It is equivalent to the one constructed in~\cite{Campoamor-Stursberg2005b},
but it contains no complex numbers and is written in a more compact form.

\section{Notations}\label{SectionNotations}

Further we use the following notations: 

$\diag(\alpha_1,\ldots,\alpha_k)$ is the $k\times k$ diagonal matrix with the elements $\alpha_1$, \ldots, $\alpha_k$ 
on the diagonal;

$E^k=\diag(1,\ldots,1)$ is the $k\times k$ unity matrix;

$E^k_{ij}$ (for the fixed values $i$ and $j$) denotes the $k\times k$ matrix $(\delta_{ii'}\delta_{jj'})$ 
with $i'$ and $j'$ running the numbers of rows and column correspondingly, 
i.e. the $k\times k$ matrix with the unit on the cross of the $i$-th
row and the $j$-th column and the zero otherwise;

$J_\lambda^k$ is the Jordan block of dimension~$k$ and the eigenvalue $\lambda$: 
\[
[J_\lambda^k]_{ij}=\left\{
\begin{array}{ll}
\lambda, & \mbox{if}\ j=i, \\
1, & \mbox{if}\ j-i=1, \\
0, & \mbox{otherwise,}
\end{array}%
\right. \quad i,j=1,\ldots,k, 
\]
i.e.
\[
J_\lambda^k={\arraycolsep=0ex
\left(\begin {array}{cccccc}
\lambda & 1 & 0 & 0 & \cdots & 0 \\
0 & \lambda & 1 & 0 & \cdots & 0 \\
0 & 0 & \lambda & 1 & \cdots & 0 \\
\cdots & \cdots & \cdots & \cdots & \cdots & \cdots\\ 
0 & 0 & 0 & 0 & \cdots & 1 \\
0 & 0 & 0 & 0 & \cdots & \lambda
\end {array}\right)},\quad
\exp(\theta J_0^k)={
\left(\begin {array}{cccccc}
1 & \theta& \frac{1}{2!}\theta^2 & \frac{1}{3!}\theta^3 & \cdots & \frac{1}{(k-1)!}\theta^{k-1}\\
0 & 1 & \theta& \frac{1}{2!}\theta^2 & \cdots & \frac{1}{(k-2)!}\theta^{k-2}\\
0 & 0 & 1 & \theta& \cdots & \frac{1}{(k-3)!}\theta^{k-3}\\
\cdots & \cdots & \cdots & \cdots & \cdots & \cdots\\ 
0 & 0 & 0 & 0 & \cdots & \theta\\
0 & 0 & 0 & 0 & \cdots & 1
\end {array}\right)}
\]
(let us note that $J_\lambda^k=\lambda E^k+J_0^k$ 
and therefore $\exp(\theta J_\lambda^k)=e^{\lambda\theta}\exp(\theta J_0^k)$\,);

$R_{\mu\nu}^r$ is the real Jordan block of dimension~$r=2k$, $k\in\mathbb{N}$, which corresponds to 
the pair of two complex Jordan blocks $J_\lambda^k$ and $J_{\lambda^*}^k$ with 
the complex conjugate eigenvalues $\lambda$ and $\lambda^*$, where 
$\mu=\mathop{\rm Re}\lambda$, $\nu=\mathop{\rm Im}\lambda\not=0$:
\[
R_{\mu\nu}^2=\left(\begin {array}{cc}
\mu & \nu \\
-\nu & \mu 
\end {array}\right),\quad
R_{\mu\nu}^{2k}=\left.
\left(\begin {array}{cccccc}
R_{\mu\nu}^2 & E^2 & 0 & 0 & \cdots & 0 \\
0 & R_{\mu\nu}^2 & E^2 & 0 & \cdots & 0 \\
0 & 0 & R_{\mu\nu}^2 & E^2 & \cdots & 0 \\
\cdots & \cdots & \cdots & \cdots & \cdots & \cdots\\ 
0 & 0 & 0 & 0 & \cdots & E^2 \\
0 & 0 & 0 & 0 & \cdots & R_{\mu\nu}^2
\end {array}\right)\right\}k\ \mbox{blocks};
\]

$A_1\oplus A_2$ is the direct sum $\arraycolsep=0.5ex\left(\begin{array}{cc}A_1&0\\0&A_2\end{array}\right)$ 
of the square matrices $A_1$ and $A_2$; 

$A_1\stackrel{C}{+}A_2$ is the block triangular matrix 
$\arraycolsep=0.5ex\left(\begin{array}{cc}A_1&C\\0&A_2\end{array}\right)$, 
where $A_1\in M_{k,k}$, $A_2\in M_{l,l}$, $C\in M_{k,l}$. 

Above 0 denotes the zero matrices of different dimensions.

\section{Solvable algebras with Abelian ideals of codimension 1}\label{SectionOnAlgsWithAbIdialsOfCodim1}

Consider a Lie algebra~$\mathfrak{g}$ of dimension~$n$ with the Abelian ideal~$I$ of dimension~$n-1$ 
(cf.~\cite{Mubarakzyanov1963-1}).
Let us suppose that the ideal~$I$ is spanned on the basis elements~$e_1$, $e_2$, \ldots, $e_{n-1}$.
Then the algebra~$\mathfrak{g}$ is completely determined by the $(n-1)\times(n-1)$ matrix~$M=(m_{kl})$ 
of restriction of the adjoint action ${\rm ad}_{e_n}$ on the ideal~$I$. 
The (possibly) non-zero commutation relations of~$\mathfrak{g}$
have the form 
\begin{gather*}
[e_k,e_n]=\sum_{l=1}^{n-1}m_{lk}e_l, \qquad k =1,\ldots,n-1.
\end{gather*}
Due to possibility of scaling~$e_{n}$, the matrix~$M$ and, therefore, its eigenvalues are determined 
up to multiplication on a non-zero number from the field under consideration. 
The matrix~$M$ is reduced to the Jordan canonical form by change of the basis in~$I$: 
\[
M=J_{\lambda_1}^{r_1}\oplus \cdots \oplus J_{\lambda_s}^{r_s},
\]
where $r_1+\cdots +r_s=n-1$, $r_i\in\mathbb{N}$, $\lambda_i\in {\mathbb C}$, $i=1,\ldots,s$. 
In the real case the direct sum of two complex blocks $J_{\lambda_i}^{r_i}$ and $J_{\lambda_j}^{r_j}$, 
where $r_i=r_j$ and $\lambda_i$ is conjugate of $\lambda_j$, is assumed as replaced 
by the corresponding real Jordan block~$R_{\mu\nu}^{2r_i}$ 
with $\mu=\mathop{\rm Re}\lambda_i$ and $\nu=\mathop{\rm Im}\lambda_i\not=0$.
The Jordan canonical form is unique up to permutation of the Jordan blocks.

The above algebra will be denoted as ${\mathfrak J}_{\lambda_1\ldots\lambda_s}^{r_1\ldots r_s}$. 
It is additionally assumed that ${\mathfrak J}_{\lambda'_1\ldots\lambda'_{s'}}^{r'_1\ldots r'_{s'}}$ 
denotes the same algebra if $s=s'$ and there exists a non-zero constant~$\varkappa$ that
$(\lambda'_i,r'_i)=(\varkappa\lambda_i,r_i)$, $i=1,\ldots,s$, up to permutation of pairs $(\lambda_i,r_i)$.

The Lie algebra ${\mathfrak J}_{\lambda_1\ldots\lambda_s}^{r_1\ldots r_s}$ 
is decomposable iff there exists a value of~$i$ such that $(\lambda_i,r_i)=(0,1)$. 
(Then $e_i$ is an invariant of ${\mathfrak J}_{\lambda_1\ldots\lambda_s}^{r_1\ldots r_s}$.)
Hence the contrary condition is supposed to be satisfied below. 
It should be also noted this algebra is nilpotent iff $\lambda_1=\cdots=\lambda_s=0$.

\subsection{Simplest cases}

Consider the simplest case for $M$ to be a single Jordan block with the eigenvalue $\lambda$, 
i.e.\ $\mathfrak{g}={\mathfrak J}_\lambda^{n-1}$, $n=2,4,\ldots$\,.
The value of~$\lambda$ can be normalized to 1 in case $\lambda\not=0$ 
but it is convenient for the further consideration to avoid normalization of~$\lambda$ some time. 

The non-zero commutation relations of~${\mathfrak J}_\lambda^{n-1}$ at most are 
\begin{gather*}
[e_1,e_n]=\lambda e_{1}, \qquad [e_k,e_n]=\lambda e_{k}+e_{k-1}, \quad k=2,\ldots,n-1, \quad \lambda \in{\mathbb C}.
\end{gather*}
(The first one is zero if $\lambda=0$.)
Therefore, its inner automorphisms are described by the triangular matrix 
\[
B(\theta)=\exp(\theta_n J^{n-1}_\lambda)\stackrel{C}{+}E^1, \quad 
C=(\theta_2+\lambda \theta_1,\theta_3+\lambda \theta_2,\ldots,
\theta_{n-1}+\lambda \theta_{n-2},\lambda \theta_{n-1})^{\rm T},
\]
i.e.\ a functional basis of lifted invariants is formed by
\begin{gather}
\widehat{\mathcal{I}}_k=e^{\lambda \theta_n}\mathcal{I}_k, \quad k=1,\ldots,n-1,\qquad
\widehat{\mathcal{I}}_n=\mathcal{I}_n+\lambda\sum_{j=1}^{n-1} \theta_{j}x_j, \nonumber\\
\intertext{where} 
\label{EqInvsOfNn1}
\mathcal{I}_k=\sum_{j=1}^k \frac{\theta_n^{k-j}}{(k-j)!} x_j, \quad k=1,\ldots,n-1,\qquad
\mathcal{I}_n=\sum_{j=1}^{n-2} \theta_{j+1} x_{j}+x_n.
\end{gather}

The nilpotent ($\lambda=0$) and solvable ($\lambda\not=0$) cases of~${\mathfrak J}_\lambda^{n-1}$ 
should be considered further separately since there exists difference in the normalization procedure.
The dimension $n=2$ is singular in both  the cases. 
${\mathfrak J}_0^1$ is the two-dimensional Abelian Lie algebra and therefore has two independent invariants, 
namely $e_1$ and $e_2$. 
${\mathfrak J}_1^1$ is the two-dimensional non-Abelian Lie algebra and therefore has no invariants. 
We assume below that $n\geqslant3$.

The algebra~${\mathfrak J}_0^{n-1}$ is, in some sense, the simplest filiform algebra of dimension~$n$. 
Let us note that the adjoint representation of~${\mathfrak J}_0^{n-1}$ is unfaithful 
since the center $Z({\mathfrak J}_0^{n-1})=\langle e_1\rangle\not=\{0\}$. 
Therefore, there are $n-1$ parameters in the expression of~$B(\theta)$ excluding~$\theta_1$, 
and $\widehat{\mathcal{I}}$ coincides with~$\mathcal{I}$.  
It is obvious that the element~$e_1$ generating $Z({\mathfrak J}_0^{n-1})$ 
is one of the invariants, which corresponds to $\mathcal{I}_1=x_1$.
Another $(n-3)$ invariants are found by the normalization procedure applied 
to the lifted invariants~$\mathcal{I}_2$, \ldots, $\mathcal{I}_{n-1}$. 
Namely, we solve the equation $\mathcal{I}_2=0$ with respect to $\theta_n$ and then 
substitute the obtained expression $\theta_n=-x_2/x_1$ into the other~$\mathcal{I}$'s. 
To construct polynomial invariants finally, we multiply the derived invariants by powers 
of the invariant~$x_1$. Since the symmetrization procedure is trivial for this algebra, 
we get the following complete set of independent generalized Casimir operators 
which are classical (i.e.\ polynomial) Casimir operators:
\begin{gather}\label{InvariantsOfNn1}
\xi_1=e_1, \quad \xi_k=\sum_{j=1}^k \frac{(-1)^{k-j}}{(k-j)!}e_1^{j-2}e_2^{k-j} e_j, \quad k=3,\ldots,n-1.
\end{gather}
This set was first constructed by the moving frame approach in Example~6 of \cite{Boyko&Patera&Popovych2006} 
and completely coincides with the one determined in Lemma 1 of 
\cite{Ndogmo&Wintenitz1994b} and Theorem~4 of~\cite{Snobl&Winternitz2005}.

In case $\lambda \not=0$ the $n-2$ invariants of~${\mathfrak J}_\lambda^{n-1}$ 
are found by the normalization procedure applied 
to the lifted invariants~$\widehat{\mathcal{I}}_1$, \ldots, $\widehat{\mathcal{I}}_{n-1}$. 
We solve $\widehat{\mathcal{I}}_2=0$ with respect to the parameter $\theta_n$. 
Substitution of the obtained expression $\theta_n=-x_2/x_1$ 
into~$\widehat{\mathcal{I}}_1$ and $\widehat{\mathcal{I}}_k/\widehat{\mathcal{I}}_1$, $k=3,\ldots,n-1$,
results in a basis of~$\mathop{\rm Inv}\nolimits({\mathfrak J}_\lambda^{n-1})$:
\begin{gather*}
\zeta_1=e_1\exp\left(-\lambda \frac{e_2}{e_1}\right), \quad \zeta_k=\dfrac{\xi_k}{\xi_1^{k-1}} , \quad k=3,\ldots,n-1,
\end{gather*}
where $\xi_k$, $k=1,3,\ldots,n-1$, are defined by \eqref{InvariantsOfNn1}.

This set of invariants completely 
coincides with the one determined in Lemma 2 of \cite{Ndogmo&Wintenitz1994b}. 
We only use exponential function instead of the logarithmic one in the expression of the first invariant. 
Let us emphasize that any basis of $\mathop{\rm Inv}\nolimits({\mathfrak J}_\lambda^{n-1})$ contains 
at least one transcendental invariant. 
The other basis invariants can be chosen rational. 

The real version~${\mathfrak J}_{(\mu,\nu)}^{n-1}$ 
of the complex algebra~${\mathfrak J}_{\lambda\lambda^*}^{r\,r}$, 
where $n=2r+1$, $r\in \mathbb{N}$, $\mu=\mathop{\rm Re}\lambda$, $\nu=\mathop{\rm Im}\lambda\not=0$, 
has the non-zero commutation relations 
\begin{gather*}
[e_1,e_n]=\mu e_1-\nu e_2, \quad [e_2,e_n]=\nu e_1+\mu e_2,\\
[e_{2k-1},e_n]=\mu e_{2k-1}-\nu e_{2k}+e_{2k-3}, \quad 
[e_{2k},e_n]=\nu e_{2k-1}+\mu e_{2k}+e_{2k-2}, 
\quad k=2,\ldots,r.
\end{gather*}
A complete tuple $\widehat{\mathcal{I}}$ of lifted invariants has the form
\begin{gather*}
\widehat{\mathcal{I}}_{2k-1}=e^{\mu\theta_n}(\mathcal{I}_{2k-1}\cos\nu\theta_n-\mathcal{I}_{2k}\sin\nu\theta_n),\quad
\widehat{\mathcal{I}}_{2k}=e^{\mu\theta_n}(\mathcal{I}_{2k-1}\sin\nu\theta_n+\mathcal{I}_{2k}\cos\nu\theta_n),
\\
\widehat{\mathcal{I}}_n=\sum_{j=1}^r\Bigl(\theta_{2j-1}(\mu x_{2j-1}-\nu x_{2j})
+\theta_{2j}(\nu x_{2j-1}+\mu x_{2j})\Bigr)+
\sum_{j=1}^{r-1}\Bigl(\theta_{2j+1}x_{2j-1}+\theta_{2j+2}x_{2j}\Bigr)+x_n,
\end{gather*}
where $k=1,\ldots,r$,
\begin{gather*}
\mathcal{I}_{2k-1}=\sum_{j=1}^k\frac{\theta_n^{k-j}}{(k-j)!} x_{2j-1},\quad
\mathcal{I}_{2k}=\sum_{j=1}^k\frac{\theta_n^{k-j}}{(k-j)!} x_{2j}.
\end{gather*}

The normalization procedure is conveniently applied to the following combinations of the lifted invariants
$\widehat{\mathcal{I}}_{2k-1}$, $\widehat{\mathcal{I}}_{2k}$, $k=1,\ldots,r$:
\begin{gather*}
\widehat{\mathcal{I}}_1^2+\widehat{\mathcal{I}}_2^2=(x_1^2+x_2^2)e^{2\mu\theta_n},
\quad
\arctan\frac{\widehat{\mathcal{I}}_2}{\widehat{\mathcal{I}}_1}=\arctan\frac{x_2}{x_1}+\nu\theta_n,
\\[1ex]
\frac{\widehat{\mathcal{I}}_1\widehat{\mathcal{I}}_3+\widehat{\mathcal{I}}_2\widehat{\mathcal{I}}_4}%
{\widehat{\mathcal{I}}_1^2+\widehat{\mathcal{I}}_2^2}=
\frac{x_1x_3+x_2x_4}{x_1^2+x_2^2}+\theta_n,
\quad
\frac{\widehat{\mathcal{I}}_2\widehat{\mathcal{I}}_3-\widehat{\mathcal{I}}_1\widehat{\mathcal{I}}_4}%
{\widehat{\mathcal{I}}_1^2+\widehat{\mathcal{I}}_2^2}=
\frac{x_2x_3-x_1x_4}{x_1^2+x_2^2},
\\[1ex]
\frac{\widehat{\mathcal{I}}_1\widehat{\mathcal{I}}_{2k-1}+\widehat{\mathcal{I}}_2\widehat{\mathcal{I}}_{2k}}%
{\widehat{\mathcal{I}}_1^2+\widehat{\mathcal{I}}_2^2}=
\frac{x_1\mathcal{I}_{2k-1}+x_2\mathcal{I}_{2k}}{x_1^2+x_2^2},
\quad
\frac{\widehat{\mathcal{I}}_2\widehat{\mathcal{I}}_{2k-1}-\widehat{\mathcal{I}}_1\widehat{\mathcal{I}}_{2k}}%
{\widehat{\mathcal{I}}_1^2+\widehat{\mathcal{I}}_2^2}=
\frac{x_2\mathcal{I}_{2k-1}-x_1\mathcal{I}_{2k}}{x_1^2+x_2^2},
\quad 
k=3,\ldots,r.
\end{gather*}
We use the condition that the third combination 
(or second one if $n=3$) equals to 0 as a normalization equation on the parameter $\theta_n$  
and then exclude $\theta_n$ from the other combinations. 
It gives the basis of~$\mathop{\rm Inv}\nolimits({\mathfrak J}_{(\mu,\nu)}^{n-1})$   
\begin{gather*}
\zeta_1=(e_1^2+e_2^2)\exp\Bigl(-2\frac\mu\nu\arctan\frac{e_2}{e_1}\Bigr),
\\[1ex]
\zeta_3=\nu\frac{e_1e_3+e_2e_4}{e_1^2+e_2^2}-\arctan\frac{e_2}{e_1},
\quad
\zeta_4=\frac{e_1e_4-e_2e_3}{e_1^2+e_2^2},
\\[1ex]
\zeta_{2k-1}=\frac{e_1\hat\zeta_{2k-1}+e_2\hat\zeta_{2k}}{e_1^2+e_2^2}, \quad 
\zeta_{2k}=\frac{e_2\hat\zeta_{2k-1}-e_1\hat\zeta_{2k}}{e_1^2+e_2^2},\quad k=3,\ldots,r,
\end{gather*}
where 
\[
\hat\zeta_{2k-1}=\sum_{j=1}^k
\left(-\frac{e_1e_3+e_2e_4}{e_1^2+e_2^2}\right)^{k-j}\frac{e_{2j-1}}{(k-j)!},\qquad
\hat\zeta_{2k}=\sum_{j=1}^k
\left(-\frac{e_1e_3+e_2e_4}{e_1^2+e_2^2}\right)^{k-j}\frac{e_{2j}}{(k-j)!}.
 \]
Therefore, ${\mathfrak J}_{(\mu,\nu)}^{2}$ has unique independent invariant~$\zeta_1$ which is necessarily transcendental.
In case $n=2r+1\geqslant5$ any basis of~$\mathop{\rm Inv}\nolimits({\mathfrak J}_{(\mu,\nu)}^{n-1})$ contains at least 
two transcendental invariants; the other $n-4$ basis invariants can be chosen rational. 
A quite optimal basis with minimal number of transcendental invariants is formed by $\zeta_k$, $k=1,3,\ldots,n-1$.

\subsection{General case}\label{SubsecOnGeneralJordanForm}

The inner automorphisms of~${\mathfrak J}_{\lambda_1\ldots\lambda_s}^{r_1\ldots r_s}$ 
are described by the triangular matrix 
\begin{gather*}
B(\theta)=\bigl(\exp(\theta_n J^{r_1}_{\lambda_1})\oplus \cdots \oplus \exp(\theta_n J^{r_s}_{\lambda _s})\bigr)
\stackrel{C}{+}E^1, \quad
C=(C^{r_1}_{\lambda_1},\ldots,C^{r_s}_{\lambda_s})^{\rm T},\\
C_{\lambda_j}^{r_j}=(
\lambda_j\theta_{\rho_j+1}+\theta_{\rho_j+2},\,\ldots,\,
\lambda_j\theta_{\rho_j+r_j-1}+\theta_{\rho_j+r_j},\, 
\lambda_j\theta_{\rho_j+r_j}), \quad j=1,\ldots,s, \\ 
\rho_1=0,\quad \rho_j=r_1+\cdots+r_{j-1}, \quad j=2,\ldots,s.
\end{gather*}
The corresponding complete tuple $\widehat{\mathcal{I}}=\check x\cdot B(\theta)$ of lifted invariants has the form
\begin{gather*}
\widehat{\mathcal{I}}_{\rho_j+q} =e^{\lambda_j\theta_n}\sum_{p=1}^q \frac{1}{(q-p)!}\theta_n^{q-p} x_{\rho_j+p},
\quad j=1,\ldots,s, \quad q =1,\ldots,r_j,\\
\widehat{\mathcal{I}}_n=\sum_{j=1}^s\biggl(\lambda_j\sum_{q=1}^{r_j}\theta_{\rho_j+q}x_{\rho_j+q}+
\sum_{q=1}^{r_j-1}\theta_{\rho_j+q+1}x_{\rho_j+q}\biggr)+x_n.
\end{gather*}
This tuple is obviously modified in the real case with complex eigenvalues.  

The $n-2$ invariants are found by the normalization procedure applied 
to the lifted invariants $\widehat{\mathcal{I}}_1$,~\ldots, $\widehat{\mathcal{I}}_{n-1}$ in different ways. 
We can either use the same normalization equation for all Jordan blocks or 
normalize lifted invariants for each Jordan block separately and then simultaneously normalize 
some lifted invariants corresponding to different Jordan blocks. 
Intermediate variants are also possible. 
In any case, the procedure is reduced to choice of $n-2$ pairs 
from the lifted invariants~$\widehat{\mathcal{I}}_1$,~\ldots, $\widehat{\mathcal{I}}_{n-1}$. 
The first term of each pair gives the left-hand side of the corresponding normalization equations. 
Substitution of the obtained value of the parameter~$\theta_n$ into the second term of the pair results in 
an invariant of~${\mathfrak J}_{\lambda_1\ldots\lambda_s}^{r_1\ldots r_s}$.
The constructed invariants form 
a basis of~$\mathop{\rm Inv}\nolimits({\mathfrak J}_{\lambda_1\ldots\lambda_s}^{r_1\ldots r_s})$
iff each from the lifted invariants $\widehat{\mathcal{I}}_1$,~\ldots, $\widehat{\mathcal{I}}_{n-1}$ 
falls within the $n-2$ chosen pairs at least once.   

We use the strategy based on initial normalization of lifted invariants for each Jordan block separately.
Then it is sufficient for us to describe the procedure for different kinds of pairs of Jordan blocks. 
Below we adduce short explanation on these pairs together with the optimally used pairs of lifted invariants 
and obtained invariants of the algebra; $i,j=1,\ldots,s$.
\begin{gather*}
J_{\lambda_i}^{r_i},\ J_{\lambda_j}^{r_{\smash{j}\vphantom{i}}}\colon
\\[1ex] 
\lambda_i\not=0,\ \lambda_j\not=0\colon\quad 
\widehat{\mathcal{I}}_{\rho_i+1},\ \widehat{\mathcal{I}}_{\rho_j+1}, \quad
e_{\rho_i+1}^{-\lambda_{\smash{j}\vphantom{i}}}e_{\rho_j+1}^{\lambda_i};
\\[1ex]
r_i\geqslant2,\ \lambda_i=0,\ r_j\geqslant2,\ \lambda_j=0\colon\quad 
\widehat{\mathcal{I}}_{\rho_i+2},\ \widehat{\mathcal{I}}_{\rho_j+2}, \quad
e_{\rho_j+2}e_{\rho_i+1}-e_{\rho_i+2}e_{\rho_j+1};
\\[1ex]
r_i\geqslant2,\ \lambda_i\not=0,\ r_j\geqslant2,\ \lambda_j=0\colon\quad 
\widehat{\mathcal{I}}_{\rho_i+2},\ \widehat{\mathcal{I}}_{\rho_j+2}, \quad
\frac{e_{\rho_j+2}}{e_{\rho_j+1}}-\frac{e_{\rho_i+2}}{e_{\rho_i+1}};
\\
r_i=1,\ \lambda_i\not=0,\ r_j\geqslant2,\ \lambda_j=0\colon\quad 
\widehat{\mathcal{I}}_{\rho_i+1},\ \widehat{\mathcal{I}}_{\rho_j+2}, \quad
e_{\rho_i+1}\exp\Bigl(-\lambda_i\frac{e_{\rho_j+2}}{e_{\rho_j+1}}\Bigr);
\\[1ex]
J_{\lambda_i}^{r_i},\ R_{\mu_j\nu_j}^{2r_{\smash{j}\vphantom{i}}},\ 
\lambda_i,\mu_j,\nu_j\in\mathbb{R},\ \nu_j\not=0\colon
\\[1ex] 
r_i\geqslant2,\ r_j\geqslant2\colon\quad 
\widehat{\mathcal{I}}_{\rho_i+2},\ 
\frac{\widehat{\mathcal{I}}_{\rho_j+2}\widehat{\mathcal{I}}_{\rho_j+3}-
\widehat{\mathcal{I}}_{\rho_j+1}\widehat{\mathcal{I}}_{\rho_j+4}}%
{\widehat{\mathcal{I}}_{\rho_j+1}^2+\widehat{\mathcal{I}}_{\rho_j+2}^2}, 
\quad
\frac{e_{\rho_j+1}e_{\rho_j+3}+e_{\rho_j+2}e_{\rho_j+4}}{e_{\rho_j+1}^2+e_{\rho_j+2}^2}
-\frac{e_{\rho_i+2}}{e_{\rho_i+1}};
\\[1ex]
r_i=1\ \mbox{or}\ r_j=1\colon\quad 
\widehat{\mathcal{I}}_{\rho_i+1},\ \arctan\frac{\widehat{\mathcal{I}}_{\rho_j+2}}{\widehat{\mathcal{I}}_{\rho_j+1}},
\quad
e_{\rho_i+1}\exp\Bigl(-\frac{\lambda_i}{\nu_j}\arctan\frac{e_{\rho_j+2}}{e_{\rho_j+1}}\Bigr);
\\[1ex]
R_{\mu_i\nu_i}^{2r_i},\ R_{\mu_j\nu_j}^{2r_{\smash{j}\vphantom{i}}},\ 
\mu_i,\nu_i,\mu_j,\nu_j\in\mathbb{R},\ \nu_i\nu_j\not=0\colon
\\[1ex] 
r_i\geqslant2,\ r_j\geqslant2\colon\quad 
\frac{\widehat{\mathcal{I}}_{\rho_i+2}\widehat{\mathcal{I}}_{\rho_i+3}-
\widehat{\mathcal{I}}_{\rho_i+1}\widehat{\mathcal{I}}_{\rho_i+4}}%
{\widehat{\mathcal{I}}_{\rho_i+1}^2+\widehat{\mathcal{I}}_{\rho_i+2}^2}, 
\ 
\frac{\widehat{\mathcal{I}}_{\rho_j+2}\widehat{\mathcal{I}}_{\rho_j+3}-
\widehat{\mathcal{I}}_{\rho_j+1}\widehat{\mathcal{I}}_{\rho_j+4}}%
{\widehat{\mathcal{I}}_{\rho_j+1}^2+\widehat{\mathcal{I}}_{\rho_j+2}^2}, 
\\[1ex]
\frac{e_{\rho_j+1}e_{\rho_j+3}+e_{\rho_j+2}e_{\rho_j+4}}{e_{\rho_j+1}^2+e_{\rho_j+2}^2}-
\frac{e_{\rho_i+1}e_{\rho_i+3}+e_{\rho_i+2}e_{\rho_i+4}}{e_{\rho_i+1}^2+e_{\rho_i+2}^2};
\\[1ex]
r_i=1\ \mbox{or}\ r_j=1\colon\quad 
\arctan\frac{\widehat{\mathcal{I}}_{\rho_i+2}}{\widehat{\mathcal{I}}_{\rho_i+1}},\ 
\arctan\frac{\widehat{\mathcal{I}}_{\rho_j+2}}{\widehat{\mathcal{I}}_{\rho_j+1}},
\quad
{\nu_i}\arctan\frac{e_{\rho_j+2}}{e_{\rho_j+1}}-{\nu_j}\arctan\frac{e_{\rho_i+2}}{e_{\rho_i+1}}.
\end{gather*}

The marginal case is $r_1=\cdots=r_s=1$,
i.e.\ all Jordan blocks are one-dimensional. 
Let us recall that $\lambda_k$ is assumed non-zero if $r_k=1$. 
A complete set of generalized Casimir operators is formed by 
$e_1^{-\lambda_j}e_j^{\lambda_1}$, $j=2,\ldots,n-1$.
In case $\lambda_j/\lambda_1\in{\mathbb Q}$, the invariants can be made rational with raising to a power.
If additionally $\lambda_j/\lambda_1$ have the same sign, 
$\mathop{\rm Inv}\nolimits({\mathfrak J}_{\lambda_1\ldots\lambda_s}^{r_1\ldots r_s})$
has a polynomial basis, i.e.\ a basis consisting of usual Casimir operators.

Therefore, $\mathop{\rm Inv}\nolimits({\mathfrak J}_{\lambda_1\ldots\lambda_s}^{r_1\ldots r_s})$
has a polynomial basis only in two cases
\begin{enumerate}
\item[1)] 
$\lambda_1=\cdots=\lambda_s=0$, i.e.\ the algebra is nilpotent;
\item [2)]
$s=n-1>2$, $r_j=1$, $\lambda_j/\lambda_1$ are rational and  have the same sign, $j=2,\ldots,n-1$.
\end{enumerate}

\section{Solvable Lie algebras with nilradical {\mathversion{bold}${\mathfrak J}_0^{n-1}$}}%
\label{SectionOnAlgsWithSimplestFiliformNilradicals}

Let us pass to complex indecomposable solvable Lie algebras with the nilradicals isomorphic
to~${\mathfrak J}_0^{n-1}$, $n=4,5,\ldots\,$. 
All possible types of such algebras are described in Theorems~1--3 of~\cite{Snobl&Winternitz2005}.
Their dimensions can be equal to $n+1$ or $n+2$. 
Below we adduce only the non-zero commutation relations, excluding ones between basis elements of the nilradicals:
\begin{gather*}
[e_k,e_n]=e_{k-1}, \quad k=2,\ldots,n-1.
\end{gather*}

There exist three inequivalent classes of such algebras of dimension $n+1$.
The first series ${\mathfrak s}_{1,n+1}$ is formed by Lie algebras ${\mathfrak s}_{1,n+1}^{\alpha\beta}$
with the additional non-zero commutation relations 
\begin{gather*}
[e_k,e_{n+1}]=\gamma_k e_k, \quad k=1,\ldots,n-1, \quad [e_n,e_{n+1}]=\alpha e_n.
\end{gather*}
where $\gamma_k:=(n-k-1)\alpha+\beta$. 
Due to scale transformations of $e_{n+1}$ the parameter tuple $(\alpha,\beta)$ can be normalized 
to belong to the set $\{(1,\beta),(0,1)\}$. 
We assume below that the parameters take only the normalized values. 
Then any algebras in the series ${\mathfrak s}_{1,n+1}$ are inequivalent each to other. 
For the values $(\alpha,\beta) \in\{(1,0),(1,2-n),(0,1)\}$ the corresponding algebras have some singular properties.  

The second class consists of the unique algebra ${\mathfrak s}_{2,n+1}$:
\begin{gather*}
[e_k,e_{n+1}]=\gamma_k e_k, \quad k=1,\ldots,n-1, \quad [e_n,e_{n+1}]=e_n+e_{n+1},
\end{gather*}
where $\gamma_k:=n-k$.

The Lie algebra ${\mathfrak s}_{3,n+1}^{a_3,\ldots,a_{n-1}}$ 
from the latter $(n-3)$-parametric series ${\mathfrak s}_{3,n+1}$ 
is determined by the commutation relations 
\begin{gather*}
[e_k,e_{n+1}]=e_k+\sum_{i=1}^{k-2} a_{k-i+1}e_i, \quad k=1,\ldots,n-1,
\end{gather*}
where $a_j\in \mathbb C$, $j=3,\ldots,n-1$, and $a_j\not=0$ for some values of~$j$. 

The unique $(n+2)$-dimensional algebra ${\mathfrak s}_{4,n+2}$ of such type 
has the additional non-zero commutation relations 
\begin{gather*}
[e_k,e_{n+1}]=\gamma_k e_k,\quad [e_n,e_{n+1}]=e_n, \quad [e_k,e_{n+2}]=e_k, \quad k=1,\ldots,n-1,
\end{gather*}
where $\gamma_k:=n-k-1$.

The matrices determining the inner automorphisms of the above algebras are conveniently presented in the form 
$B(\theta)=B_1B_2B_3$, where
\begin{gather*}
B_1=\exp(\theta_1\hat{\rm ad}_{e_1})\cdots\exp(\theta_{n-1}\hat{\rm ad}_{e_{n-1}}), \quad 
B_2=\exp(-\theta_n\hat{\rm ad}_{e_n}), \quad
B_3=\exp(-\theta_{n+1}\hat{\rm ad}_{e_{n+1}}),
\end{gather*}
excluding the $(n+2)$-dimensional case where
$B_3=\exp(-\theta_{n+1}\hat{\rm ad}_{e_{n+1}})\exp(-\theta_{n+2}\hat{\rm ad}_{e_{n+2}}).$
The matrices $B_1$, $B_2$ and $B_3$ are written in a block form corresponding to partition 
of a basis of the algebra under consideration to 
the basis $e_1$, \ldots, $e_{n-1}$ of the maximal Abelian ideal and a complementary part. 

{\samepage
For the algebra~${\mathfrak s}_{1,n+1}^{\alpha\beta}$
\begin{gather*}
B_1=E^{n-1}\stackrel{C}{+}E^2, 
\\[1ex]
B_2=\exp(\theta_nJ_0^{n-1})\oplus
\left(\begin {array}{cc}
1 & -\alpha\theta_n\\ 0  & 1
\end {array}\right),\hspace{8em} 
C={}\raisebox{0ex}[0ex][0ex]{$\displaystyle
\left(\begin {array}{cc}
\theta_2 & \gamma_1\theta_1\\
\theta_3 & \gamma_2\theta_2\\
\cdots & \cdots\\ 
\theta_{n-1} & \gamma_{n-2}\theta_{n-2}\\
0  & \gamma_{n-1}\theta_{n-1}
\end {array}\right),$}
\\[1ex]
B_3=\diag(e^{\gamma_1\theta_{n+1}},\ldots,e^{\gamma_{n-1}\theta_{n+1}})\oplus\diag(e^{\alpha\theta_{n+1}},1).
\end{gather*}}

\noindent
Therefore, the corresponding complete tuple $\widehat{\mathcal{I}}=\check x\cdot B(\theta)$ of lifted invariants has the form
\begin{gather*}
\widehat{\mathcal{I}}_k=e^{\gamma_k\theta_{n+1}}\mathcal{I}_k, \quad k=1,\ldots,n-1,\\
\widehat{\mathcal{I}}_n=e^{\alpha \theta_{n+1}}\mathcal{I}_n,\quad 
\widehat{\mathcal{I}}_{n+1}=-\alpha \theta_n \mathcal{I}_n+\mathcal{I}_{n+1}.
\end{gather*}
Here $\mathcal{I}_1$, \ldots, $\mathcal{I}_n$ are defined by~\eqref{EqInvsOfNn1}, 
and 
\[
\mathcal{I}_{n+1}:=\sum_{j=1}^{n-1} \gamma_j\theta_{j}x_j+ x_{n+1}.
\]
 
For the algebra ${\mathfrak s}_{2,n+1}$ only the matrices $B_2$ and $B_3$ 
and the lifted invariants~$\widehat{\mathcal{I}}_n$ and~$\widehat{\mathcal{I}}_{n+1}$
differ from those of the previous algebras: 
\begin{gather*}
B_2=\exp(\theta_nJ_0^{n-1})\oplus
\left(\begin {array}{cc}
1 & e^{-\theta_n}-1\\ 0  & e^{-\theta_n}
\end {array}\right),
\\[1ex]
B_3=\diag(e^{\gamma_1\theta_{n+1}},\ldots,e^{\gamma_{n-1}\theta_{n+1}})\oplus
\left(\begin {array}{cc}
e^{\theta_{n+1}} & 0\\
e^{\theta_{n+1}}-1 & 1
\end {array}\right),
\\[1ex]
\widehat{\mathcal{I}}_n=\big(e^{\theta_{n+1}-\theta_n}-e^{-\theta_n}+1\big)\mathcal{I}_n
+e^{-\theta_n}\big(e^{\theta_{n+1}}-1\big)\mathcal{I}_{n+1},
\quad 
\widehat{\mathcal{I}}_{n+1}=\big(e^{-\theta_n}-1\big)\mathcal{I}_n+e^{-\theta_n}\mathcal{I}_{n+1}. 
\end{gather*}

All the $(n+1)$-dimensional algebras under consideration have exactly $n-3$ independent invariants 
which can be found by the normalization procedure applied 
to the lifted invariants~$\widehat{\mathcal{I}}_1$, \ldots, $\widehat{\mathcal{I}}_{n-1}$. 
Since the invariants of these algebras depend only on element of the Abelian ideal 
the symmetrization procedure is trivial and can be omitted as a step of the algorithm.

For the algebras~${\mathfrak s}_{1,n+1}^{\alpha\beta}$, $(\alpha,\beta)\not=(1,2-n)$ and ${\mathfrak s}_{2,n+1}$
we solve equations $\widehat{\mathcal{I}}_1=1$ and  $\widehat{\mathcal{I}}_2=0$
with respect to the values $e^{\theta_{n+1}}$ and $\theta_{n}$.
Substituting the obtained expressions 
$e^{\theta_{n+1}}=x_1^{-1/\gamma_1}$ and $\theta_n=-{x_2}/{x_1}$ into the other~$\widehat{\mathcal{I}}$'s,
we get the following complete set of generalized Casimir operators
\[
\xi_1^{-(k-1)\frac{(n-3)\alpha+\beta}{(n-2)\alpha+\beta}}\xi_k, \quad k=3,\ldots,n-1, 
\]
where $\xi_k$, $k=1,3,\ldots,n-1$, are defined by \eqref{InvariantsOfNn1}.
For the algebra ${\mathfrak s}_{2,n+1}$ the value $\alpha=\beta=1$ should be taken.

The algebra ${\mathfrak s}_{1,n+1}^{1,2-n}$ is singular with respect to the normalization procedure 
and will be studied separately.
In this case $\gamma_1=0$ and $\widehat{\mathcal{I}}_1=x_1$ hence  
the basis element~$e_1$ generating the center of~${\mathfrak s}_{1,n+1}^{1,2-n}$ is one of the invariants. 
We obtain the expressions for $\theta_{n}$ and $e^{\theta_{n+1}}$ from the system 
$\widehat{\mathcal{I}}_2=0$,  $\widehat{\mathcal{I}}_3=1$ and
substitute them into the other $\widehat{\mathcal{I}}$'s. 
Additionally we use the possibility on multiplication of invariants by powers of the invariant~$x_1$.
The resulting complete set of generalized Casimir operators is formed by 
\begin{gather*}
\xi_1, \quad \frac{\xi_k^2}{\xi_3^{k-1}} , \quad k=4,\ldots,n-1.
\end{gather*}

Calculations for the Lie algebra~~${\mathfrak s}_{3,n+1}^{a_3,\ldots,a_{n-1}}$ are analogous but more complicated:
\begin{gather*}
B_1=E^{n-1}\stackrel{C}{+}E^2, 
\qquad 
C=
\left(\begin {array}{cc}
\theta_2 & \theta_1+a_3\theta_3+a_4\theta_4+\cdots+a_{n-1}\theta_{n-1}\\
\theta_3 & \theta_2+a_3\theta_4+a_4\theta_5+\cdots+a_{n-2}\theta_{n-1}\\
\cdots & \cdots\\ 
\theta_{n-3} & \theta_{n-3}+a_3\theta_{n-2}\\
\theta_{n-1} & \theta_{n-2}\\
0  & \theta_{n-1}
\end {array}\right),
\\[1ex]
B_2=\exp(\theta_nJ_0^{n-1})\oplus E^2,\\[1ex]
B_3=e^{\theta_{n+1}}
\Biggl(E^{n-1}+\sum_{m=2}^{n-2}(J_0^{n-1})^m
\sum_{i=1}^{\left[\frac m2\right]}\frac{b_{mi}}{i!}\theta^i_{n+1}\Biggr)\oplus E^2, \qquad 
b_{mi}=\sum_{\substack{3\leqslant s_1,\ldots,s_i\leqslant n-1\\s_1+\cdots+s_i=m+i}}a_{s_1}\cdots a_{s_i}.
\end{gather*}
The corresponding complete tuple $\widehat{\mathcal{I}}=\check x\cdot B(\theta)$ of lifted invariants has the form
\begin{gather*}
\widehat{\mathcal{I}}_k=e^{\theta_{n+1}}\Biggl(\mathcal{I}_k+\sum_{m=2}^{k-1}\mathcal{I}_{k-m}
\sum_{i=1}^{\left[\frac m2\right]}\frac{b_{mi}}{i!}\theta^i_{n+1}\Biggr) , \quad k=1,\ldots,n-1,
\\[1ex]
\widehat{\mathcal{I}}_n=\mathcal{I}_n,\quad 
\widehat{\mathcal{I}}_{n+1}=\mathcal{I}_{n+1}+\sum_{k=1}^{n-1}x_k\sum_{i=1}^{n-k-2} \theta_{k+1+i} a_{i+2}.
\end{gather*}

{\samepage
The Lie algebra~${\mathfrak s}_{3,n+1}^{a_3,\ldots,a_{n-1}}$ has $n-3$ invariants for any values of the parameters.
Applying the normalization procedure to $\widehat{\mathcal{I}}$,
we solve the system $\mathcal{I}_1=1$, $\mathcal{I}_2=0$ with respect to~$\theta_{n}$ and $\theta_{n+1}$. 
Substitution of the obtained expressions $\theta_n=-{x_2}/{x_1}$ and $\theta_{n+1}=-\ln x_1$ 
into the other~$\mathcal{I}$'s gives the following complete set of generalized Casimir operators
\[
\xi_1^{-k+1}\xi_k+\sum_{m=2}^{k-1}\xi_1^{-k+m+1}\xi_{k-m}
\sum_{i=1}^{\left[\frac m2\right]}\frac{b_{mi}}{i!}(-\ln \xi_1)^i, \quad k=3,\ldots,n-1, 
\]
where $\xi_k$, $k=1,3,\ldots,n-1$, are defined by \eqref{InvariantsOfNn1}.

}

For the algebra~${\mathfrak s}_{4,n+2}$ 
\begin{gather*}
B_1=E^{n-1}\stackrel{C}{+}E^3, \quad 
C=
\left(\begin {array}{ccc}
\theta_2 & \gamma_1\theta_1 & \theta_1\\
\theta_3 & \gamma_2\theta_2 & \theta_2\\
\cdots & \cdots & \cdots \\ 
\theta_{n-1} & \gamma_{n-2}\theta_{n-2} & \theta_{n-2}\\
0  & \gamma_{n-1}\theta_{n-1} & \theta_{n-1}
\end {array}\right), \quad \gamma_k:=n-k-1,
\\[1ex]
B_2=\exp(\theta_nJ_0^{n-1})\oplus
\left(\begin {array}{ccc}
1 & -\theta_n & 0\\ 0  & 1 & 0\\ 0  & 0 & 1\\
\end {array}\right),
\\[1.5ex]
B_3=e^{\theta_{n+2}}\diag(e^{\gamma_1\theta_{n+1}},\ldots,e^{\gamma_{n-1}\theta_{n+1}})\oplus
\diag(e^{\alpha\theta_{n+1}},1,1),
\end{gather*}
i.e.\ the tuple  $\widehat{\mathcal{I}}=\check x\cdot B(\theta)$ of lifted invariants has the form
\begin{gather*}
\widehat{\mathcal{I}}_k=e^{\gamma_k\theta_{n+1}+\theta_{n+2}}\mathcal{I}_k, \quad k=1,\ldots,n-1,
\\
\widehat{\mathcal{I}}_n=e^{\theta_{n+1}}\mathcal{I}_n,\quad 
\widehat{\mathcal{I}}_{n+1}=\mathcal{I}_{n+1}- \theta_n\mathcal{I}_n,\quad 
\widehat{\mathcal{I}}_{n+2}=\mathcal{I}_{n+2}:=\sum_{j=1}^{n-1} \theta_{j}x_j+ x_{n+2}.
\end{gather*}

{\samepage
The $n-4$ invariants of~${\mathfrak s}_{4,n+2}$ are found by the normalization procedure applied 
to the lifted invariants~$\widehat{\mathcal{I}}_1$, \ldots, $\widehat{\mathcal{I}}_{n-1}$. 
We solve the system $\widehat{\mathcal{I}}_1=1$, $\widehat{\mathcal{I}}_2=0$, $\widehat{\mathcal{I}}_3=1$ 
with respect to the parameters $\theta_{n}$, $\theta_{n+1}$ and  $\theta_{n+2}$ and then exclude them
from the other $\widehat{\mathcal{I}}$'s. As a result, we obtain a complete set of invariants of~${\mathfrak s}_{4,n+2}$:
\[
\dfrac{\xi_k^2}{\xi_3^{k-1}} , \quad k=4,\ldots,n-1,
\]
where $\xi_k$, $k=3,\ldots,n-1$, are defined by \eqref{InvariantsOfNn1}.}

The sets of generalized Casimir invariants for the Lie algebras
with the nilradicals isomorphic to ${\mathfrak J}_0^{n-1}$, which are
constructed in this section, coincide with the ones determined in
Theorems 5 and 6 of \cite{Snobl&Winternitz2005}.

\section{Nilpotent algebra of strictly upper triangle matrices}%
\label{SectionOnAlgebraOfStrictlyUpperTriangleMatrices}

Consider the nilpotent Lie algebra $\mathfrak{t}_0(n)$ isomorphic to the one of 
strictly upper triangle $n\times n$ matrices over the field $\mathbb F$, 
where $\mathbb F$ is either $\mathbb C$ or $\mathbb R$. 
$\mathfrak{t}_0(n)$ has dimension $n(n-1)/2$. 
It is the Lie algebra of the Lie group ${T}_0(n)$ of upper unipotent $n\times n$ matrices, 
i.e.\ upper triangular matrices with the unities on the diagonal.  

Its basis elements are convenient to enumerate with the ``increasing'' pair of indices 
similarly to the canonical basis $\{E^n_{ij},\,i<j\}$ of the isomorphic matrix algebra. 
Thus, the basis elements $e_{ij}\sim E^n_{ij}$, $i<j$, satisfy the commutation relations 
\[
[e_{ij},e_{kl}]=\delta_{jk}e_{il}-\delta_{li}e_{kj}, 
\]
where $\delta_{ij}$ is the Kronecker delta. 

Hereafter the indices $i$, $j$, $k$ and $l$ run at most from 1 to~$n$. 
Only additional constraints on the indices are indicated. 

Let $e_{ji}^*$, $x_{ji}$ and $y_{ij}$ denote 
the basis element and the coordinate function in the dual space $\mathfrak{t}_0^*(n)$ and
the coordinate function in~$\mathfrak{t}_0(n)$,
which correspond to the basis element~$e_{ij}$, $i<j$. 
We complete the sets of $x_{ji}$ and $y_{ij}$ to the matrices $X$ and $Y$ with zeros. 
Hence $X$ is a strictly lower triangle matrix and $Y$ is a strictly upper triangle one. 

\begin{lemma}\label{LemmaOnLiftedInvsOfStrictlyUpperTriangularMatrices}
A complete set of independent lifted invariants of ${\rm Ad}^*_{T_0(n)}$ 
is exhausted by the expressions
\[
\mathcal{I}_{ij}=x_{ij}+\sum_{i<i'}b_{ii'}x_{i'\!j}+\sum_{j'<j}b_{j'\!j}x_{ij'}
+\sum_{i<i'\!,\,j'<j}b_{ii'}\widehat b_{j'\!j}x_{i'\!j'}, \quad j<i,
\]
where 
$B=(b_{ij})$ is an arbitrary matrix from ${T}_0(n)$; 
$\widehat B=(\widehat b_{ij})$ is the inverse matrix of~$B$.  
\end{lemma}

\begin{proof} The adjoint action of $B\in{T}_0(n)$ on the matrix~$Y$ is 
${\rm Ad}_BY=BYB^{-1}$, i.e. 
\[
{\rm Ad}_B\sum_{i<j}y_{ij}e_{ij}=\sum_{i<j}(BYB^{-1})_{ij}e_{ij}
=\sum_{i\leqslant i'<j'\leqslant j}b_{ii'}y_{i'\!j'}\widehat b_{j'\!j}e_{ij}. 
\]
After changing $e_{ij}\to x_{ji}$, $y_{ij}\to e_{ji}^*$, $b_{ij}\leftrightarrow \widehat b_{ij}$ 
in the latter equality, we obtain the representation for the coadjoint action of~$B$
\[
{\rm Ad}_B^*\sum_{i<j}x_{ji}e_{ji}^*
=\sum_{i\leqslant i'<j'\leqslant j}b_{j'\!j}x_{ji}\widehat b_{ii'}e_{j'\!i'}^*
=\sum_{i'<j'}(BXB^{-1})_{j'\!i'}e_{j'\!i'}^*.
\]
Therefore, the elements $\mathcal{I}_{ij}$, $j<i$, of the matrix 
\[
\mathcal{I}=BXB^{-1}, \quad B\in{T}_0(n),
\]
form a complete set of independent lifted invariants of ${\rm Ad}^*_{T_0(n)}$.
\end{proof}

\begin{note}
The center of the group ${T}_0(n)$ is $Z({T}_0(n))=\{E^n+b_{1n}E^n_{1n},\ b_{1n}\in\mathbb{F}\}$.
The inner automorphism group of~$\mathfrak{t}_0(n)$ is isomorphic to the factor-group 
${T}_0(n)/Z({T}_0(n))$ and hence its dimension is $\frac12n(n-1)-1$. 
The parameter $b_{1n}$ in the above representation of lifted invariants is inessential.
\end{note}

Below $A^{i_1,i_2}_{j_1,j_2}$, where $i_1\leqslant i_2$, $j_1\leqslant j_2$,
 denotes the submatrix $(a_{ij})^{i=i_1,\ldots,i_2}_{j=j_1,\ldots,j_2}$ 
of a matrix $A=(a_{ij})$. 

\begin{lemma}\label{LemmaOnASetOfInvsOfStrictlyUpperTriangularMatrices}
A set of independent invariants of~${\rm Ad}^*_{T_0(n)}$ 
is given by the expressions
\[
\det X^{n-k+1,n}_{1,k}, \quad k=1, \ldots, \left[\frac n2\right].
\]
\end{lemma}

\begin{proof}
The derived formula for~$\mathcal{I}$ and (triangle) structure of the matrices $B$ and $X$ imply that 
\[
\mathcal{I}^{n-k+1,n}_{1,k}=
B^{n-k+1,n}_{n-k+1,n}
X^{n-k+1,n}_{1,k}
\widehat B^{1,k}_{1,k}, 
\quad k=1, \ldots, \left[\frac n2\right].
\]
(These submatrices have size $k\times k$ and lie 
in the left lower angle of~$\mathcal{I}$,  
in the right lower angle of~$B$,  
in the left lower angle of~$X$ and 
in the left upper angle of~$\widehat B$ correspondingly.)
Then 
\[
\det\mathcal{I}^{n-k+1,n}_{1,k}=\det X^{n-k+1,n}_{1,k},
\quad k=1, \ldots, \left[\frac n2\right],
\]
since $\det B^{n-k+1,n}_{n-k+1,n}=\det\widehat B^{1,k}_{1,k}=1$, 
i.e.\ $\det X^{n-k+1,n}_{1,k}$ are invariants of~${\rm Ad}^*_{T_0(n)}$  
in view of the definition of invariant. Functional independence of these invariants is obvious.
\end{proof}

\begin{lemma}\label{LemmaOnNumberOfInvsOfStrictlyUpperTriangularMatrices}
The number of independent invariants of~${\rm Ad}^*_{T_0(n)}$
is not greater than $\left[\dfrac n2\right]$.
\end{lemma}

\begin{proof}
Since $\det B=1$, $\widehat b_{ij}$ for $i<j$ is algebraic complement to~$b_{ij}$ and then
\[
\widehat b_{ij}=(-1)^{i+j}\det B^{i,j-1}_{i+1,j}=-b_{ij}+\cdots,
\]
where the rest terms are polynomial in $b_{i'\!j'}$, $i'=i,\ldots,j-1$, $j'=i+1,\ldots,j$, $(i',j')\not=(i,j)$, $i'<j'$. 
These elements $b_{i'\!j'}$ are over the leading diagonal of $B$ and not to the right of and not over~$b_{ij}$. 

We order and enumerate the lifted invariants $\mathcal{I}_{ij}$, $j<i$, $i+j\not=n+1$, in the following way:
\begin{gather*}
\mathcal{I}_{n-k+1,j},\ j=1,\ldots,\min(k-1,n-k),\quad 
\mathcal{I}_{ik},\ i=\max(k+1,n-k+2),\ldots,n,\quad \\
k=2,\ldots,n-1, 
\end{gather*}
and then enumerate them. The numeration matrix will look as
\[
\left(\begin{array}{ccccccccccc}
\times
\\m_n{-}1&\times
\\m_n{-}5&m_n{-}4&\times
\\\, m_n{-}11&\,m_n{-}10 &\ m_n{-}9 &\ \times\
\\\,\cdots&\,\cdots&\ \cdots\ &\ \cdots\ &\ \cdots\ &\ \phantom{\cdots}\
\\7&8&9&\times&\ldots&\times
\\3& 4& \times& 10& \ldots& m_n{-}6 & \times
\\1& \times& 5& 11& \ldots& m_n{-}7 & m_n{-}2& \ \times\
\\\times& 2& 6& 12& \ldots& m_n{-}8 & m_n{-}3& m_n&\ \times
\end{array}\right),
\]
where
\[
m_n=\dfrac{n(n-1)}2-\left[\dfrac n2\right].
\]
The obtain tuple of lifted invariants is denoted by $\mathcal{I}_{\prec}$. 

{\samepage
In similar way we order and enumerate the parameters $b_{ij}$, $i<j$, $i+j\not=n+1$:
\begin{gather*}
b_{n-k+1,j},\ j=\max(k+1,n-k+2),\ldots,n,\quad 
b_{ik},\ i=1,\ldots,\min(k-1,n-k),\quad \\
k=2,\ldots,n-1. 
\end{gather*}}%
The corresponding numeration matrix is obtained from the previous numeration matrix 
with transposition and inversion of order of choosing pairs from  rows and columns. 
The obtain tuple of parameters is denoted by $b_{\succ}$.

In view of the representation of lifted invariants, the Jacobian matrix $\p\mathcal{I}_{\prec}/\p b_{\succ}$ 
is block lower triangle of dimension~$m_n$ with the nonsingular blocks 
\begin{gather*}
X^{n-k+1,n}_{1,k},\ (X^{n-k+1,n}_{1,k})^{\rm T}, \ k=1, \ldots, \left[\frac n2\right],\quad 
X^{n-k+1,n}_{1,k},\ (X^{n-k+1,n}_{1,k})^{\rm T}, \ k=\left[\frac n2\right], \ldots,1,
\end{gather*}
on the leading diagonal. 
Therefore, this matrix is nonsingular and the rank of the complete Jacobian matrix 
of derivatives of the lifted invariants with respect to the parameters is not less than~$m_n$. 
Then the number of independent invariants of~$\mathfrak{t}_0(n)$ is 
\begin{gather*}
N_{\mathfrak{t}_0(n)}=\dim\mathfrak{t}_0(n)-\rank\mathfrak{t}_0(n)\leqslant \dfrac{n(n-1)}2-m_n=\left[\dfrac n2\right].
\tag*{\qed}
\end{gather*} \renewcommand{\qed}{}
\end{proof}

\begin{theorem}\label{TheoremOnBasisOfInvsOfStrictlyUpperTriangularMatrices}
A basis of~${\rm Inv}(\mathfrak{t}_0(n))$ is formed by the Casimir operators 
\[
\det(e_{ij})^{i=1,\ldots,k}_{j=n-k+1,n}, \quad k=1, \ldots, \left[\frac n2\right].
\]
\end{theorem}
\begin{proof}
Lemmas~\ref{LemmaOnASetOfInvsOfStrictlyUpperTriangularMatrices} 
and~\ref{LemmaOnNumberOfInvsOfStrictlyUpperTriangularMatrices} 
immediately result in that the expressions from Lemma~\ref{LemmaOnASetOfInvsOfStrictlyUpperTriangularMatrices} 
form a basis of~${\rm Inv}({\rm Ad}^*_{T_0(n)})$. 
Since the basis elements corresponding the coordinate functions 
in these expressions commutate to each other, the symmetrization procedure is trivial.  
\end{proof}

The above basis of invariants was first obtained in a quite heuristic way in~\cite{Tremblay&Winternitz2001}.

\section{Concluding remarks}

The algebraic algorithm for computing the invariants of Lie algebras
by means of moving frames of \cite{Boyko&Patera&Popovych2006}, intended originally for Lie
algebras of fixed relatively dimension, is shown to be an efficient
method for computing invariant operators for families of solvable
Lie algebras, which share the same structure of nilradicals, but
are  of general dimension $n<\infty$. Moreover, it is clear from
the results in this paper that the method is neither limited to
such Lie algebras nor to the problem of finding generalized Casimir
operators.

There are two other very different challenging problems in Lie
theory which we want to point out in expectation that the moving
frame method could be adapted to their solution.

Consider a pair of Lie algebras $\mathfrak{g}$ and $\mathfrak{g}'$ such that $\mathfrak{g}\supset
\mathfrak{g}'$. The generalized Casimir operators, we are finding here, clearly
stabilize $\mathfrak{g}'$ inside $\mathfrak{g}$. One may expect that there are other
functions of elements of $\mathfrak{g}$ that commute with the subalgebra $\mathfrak{g}'$.
What are they? and what is their basis? Among semisimple Lie
algebras, the answer has been given for two cases. Namely
$SU(3)\supset O(3)$ in \cite{Judd&Miller&Patera&Winternitz1974}, and $SU(4)\supset SU(2)\times
SU(2)$ in \cite{Quesne1976}.
In the first case there are two additional operators in the
universal enveloping algebra of $SU(3)$ that commutate with the
subalgebra (but do not commute among themselves!). In the second
case, four additional operators were found, two and two commuting.

The generalized Casimirs can be interpreted as a basis for the
trivial one-dimensional representation of $\mathfrak{g}$. The second problem is
to describe basis operators for other representations of $\mathfrak{g}$ than
the one-dimensional that, for example for the adjoint representation of
$\mathfrak{g}$. The answer to this question has been given for many semisimple
Lie algebras and for various their representations. See \cite{Patera2003} and
references therein.

We hope that the developed approach will be effectively used in others areas 
of mathematics and physics where the problem of finding functional bases of invariants of Lie algebras is actual.
This approach can be extended in a natural way to invariants of Lie superalgebras, Poisson--Lie algebras etc. 
Investigation of (generalized) Casimir invariants of such algebras is
an important problem of theoretical and mathematical physics, in particular, 
of the theory of integrable and superintegrable systems.

In case of low-dimensional Lie algebras our method can be easy
realized by means of symbolic calculation packages.

\bigskip

\noindent
{\bf Acknowledgments.} The work was partially supported by the National Science and Enginee\-ring
Research Council of Canada, by MITACS.
The research of R.\,P. was supported by Austrian Science Fund (FWF), Lise Meitner
project M923-N13. V.\,B. is grateful for the hospitality the Centre de
Recherches Math\'ematiques, Universit\'e de Montr\'eal.

\end{document}